\documentclass[aps,pre,twocolumn,groupedaddress]{revtex4-1}
\usepackage[dvips]{graphicx}
\usepackage{amsmath}
\usepackage{amssymb}
\usepackage{color}
\usepackage{bm}

\newcommand{\ket}[1]{|#1\rangle}

\newcommand{\brket}[3]{\langle#1|#2|#3\rangle}

\newcommand{\avg}[1]{\langle#1\rangle}
\newcommand{\tm}[1]{\textrm{#1}}

\begin{document}


\title{Entropy Production and Thermalization in the One-Atom Maser}


\author{E. Solano-Carrillo}
\affiliation{Department of Physics, Columbia University, New York, New York 10027, USA}


\begin{abstract}
In the configuration in which two-level atoms with an initial thermal distribution of their states are sent in succession to a cavity sustaining a single mode of electromagnetic radiation, one atom leaving the cavity as the next one enters it (as in the one-atom maser), Jaynes and Cummings showed that the steady state of the field, when many atoms have traversed the cavity, is thermal with a temperature different than that of the atoms in the off-resonant situation. Having an interaction between two subsystems which maintains them at different temperatures was then understood as leading to an apparent violation of energy conservation. Here we show, by calculating the quantum entropy production in the system, that this difference of temperatures is consistent with having the subsystems adiabatically insulated from each other as the steady state is approached. At resonance the insulation is removed and equilibration of the temperatures is achieved.
\end{abstract}

\pacs{}

\maketitle

\section{Introduction}
The Jaynes-Cummings model is a paradigmatic model in quantum optics \cite{JCM,ShoreKnight,Greentree}. It describes the electric dipole interaction of a two-level atom with a single mode of a quantized electromagnetic field in a cavity. In cavity QED experiments \cite{Haroche_K,Walther}, where a beam of Rydberg atoms prepared in a well-defined initial state are sent to a high-$Q$ superconducting cavity (which can be tuned to the resonance frequency of two selected neighboring levels of the atoms), the model has served to understand purely quantum phenomena such as the Rabi oscillations of the level populations, as well as the collapse-and-revival of these oscillations when the field in the cavity is a thermal field \cite{Rempe} or a coherent field \cite{Brune}. It also helps elucidating the manifestation of quantum correlations (entanglement) \cite{ShoreKnight} with possible applications in quantum information processing \cite{Ellinas}.

Despite its great success, there is still a conceptual puzzle which needs to be understood: if two-level atoms with energy splitting $\Omega$ and initial thermal distribution of their states at temperature $T_a$ are sent to a single-mode cavity with frequency $\omega$, with such a flux that the $N$th atom enters the cavity as the $(N-1)$th one leaves it, the steady state of the field is thermal at temperature  
\begin{equation}\label{Tf}
 T_f=(\omega/\Omega)\,T_a.
\end{equation}
Originally discovered by Jaynes and Cummings \cite{JCM} and realized, in principle, in a one-atom maser \cite{Meschede,Nogues,Rauschenbeutel}, the result \eqref{Tf} implies that except at resonance, when $\omega=\Omega$, the two subsystems do not equilibrate their temperatures.

This apparent violation of energy conservation, as first understood \cite{JCM}, was attributed to the neglect of the translational degrees of freedom of the atoms, which was argued to make the assumption of ``thermal'' atoms unjustified. The neglect of losses in the cavity was also considered as a possible source for the lack of temperature equilibration. 

The arguments above are, however, not entirely satisfactory, for Cummings himself in an earlier paper \cite{Cummings1} showed that a two-level atom weakly interacting with an intense black-body radiation field for a very long time, approaches a Boltzmann distribution of the two levels, irrespective of the translational motion. Also, cavities with extremely high-$Q$ values are now possible to build using superconducting materials \cite{Walther}, rendering the assumption of a lossless cavity a very good approximation. 

In this paper we show that the lack of temperature equilibration between the two subsystems can be understood by applying the theory of quantum entropy production that we have recently developed \cite{Solano2}. We show that when the steady state is approached out of resonance, the field and atoms subsystems become \emph{adiabatically insulated} from each other, with the steady state being an equilibrium state with zero total entropy production. Under this circumstance, the difference of temperatures is not puzzling since an adiabatic ``wall'' is built up, in the long run, between the subsystems.  At resonance, this insulation is not present and equilibration of the temperatures automatically takes place.

In order to make the exposition as complete as possible we organize the paper as follows: in section \ref{s2} we briefly review the important features of the Jaynes-Cummings physics which are relevant for our discussion. In section \ref{s1m} we discuss the aforementioned configuration of the one-atom maser including the derivation of \eqref{Tf}. In section \ref{s3} we calculate the entropy production in the Jaynes-Cummings thought experiment, show how it tends to its steady state value and compare it with expectations from classical nonequilibrium thermodynamics. We conclude with section \ref{s4}. 

\section{Jaynes-Cummings model}\label{s2}
The Jaynes-Cummings model describes the electric dipole coupling of a quantized single mode (energy $\omega$) of electromagnetic radiation in a cavity, with a two-level atom with energy splitting $\Omega=E_2-E_1$. The Hamiltonian describing the cavity field + atom system is, in the rotating wave approximation,
\begin{equation}
 \hat{H}=E_1\hat{\pi}\hat{\pi}^{\dagger}+E_2\hat{\pi}^{\dagger}\hat{\pi}+\omega\,\hat{a}^{\dagger}\hat{a}+\gamma\,(\hat{\pi}^{\dagger}\hat{a}+\hat{\pi} \hat{a}^{\dagger}),
\end{equation}
where $\hat{a}^{\dagger}\,(\hat{a})$ is the creation (annihilation) operator of a photon in the cavity, $\hat{\pi}^{\dagger}$ and $\hat{\pi}$ are the ladder operators for the atomic states and $\gamma$ is the strength of the dipole coupling, assumed small (see appendix \ref{tpn}): 
\begin{equation}\label{gmG}
\gamma\ll\Omega. 
\end{equation}

We work in the natural basis $\lbrace\ket{\bm{\alpha}}\rbrace=\lbrace\ket{1,n+1},\ket{2,n}\rbrace$ which diagonalizes the Hamiltonian of \emph{uncoupled} subsystems $\hat{\mathcal{H}}=E_1\hat{\pi}\hat{\pi}^{\dagger}+E_2\hat{\pi}^{\dagger}\hat{\pi}+\omega\,\hat{a}^{\dagger}\hat{a}$, where $\lbrace\ket{1},\ket{2}\rbrace$ are the two states of the atom (ground \& excited), with 
\begin{equation}
 \hat{\pi}^{\dagger}\ket{1}=\ket{2},\hspace{0.3cm}\hat{\pi}^{\dagger}\ket{2}=0,\hspace{0.6cm}\hat{\pi}\ket{2}=\ket{1},\hspace{0.3cm}\hat{\pi}\ket{1}=0,
\end{equation}
and $n$ is the number of photons in the cavity. In the sector with $n$ photons, the matrix representation $H_n$ of the total Hamiltonian then reads
\begin{equation}
H_n=
 \begin{pmatrix}
  E_1+(n+1)\omega&\gamma\sqrt{n+1}\\
\gamma\sqrt{n+1}&E_2+n\omega
 \end{pmatrix},
\end{equation}
with $H=\bigoplus_n H_n$. Defining the atomic zero of energy so that $E_1+E_2=0$, we get for the eigenvalues of $H_n$
\begin{equation}\label{epm}
 E_n^{\pm}=(n+1/2)\omega\pm\beta_n,
\end{equation}
with $2\beta_n=\sqrt{(\Omega-\omega)^2+4(n+1)\gamma^2}$ and the corresponding eigenvectors
\begin{equation}\label{phis}
 \begin{split}
  \ket{\phi_n^{+}}&=\cos\theta_n\,\ket{2,n}+\sin\theta_n\,\ket{1,n+1},\\
  \ket{\phi_n^{-}}&=-\sin\theta_n\,\ket{2,n}+\cos\theta_n\,\ket{1,n+1},
 \end{split}
\end{equation}
with $\theta_n$ indicating the angle of rotation of the basis vectors defining the $n$-photon sector, where
\begin{equation}
 \tan(2\theta_n)=\dfrac{2\gamma\sqrt{n+1}}{\Omega-\omega}.
\end{equation}
Inverting \eqref{phis} and noting that $\ket{\phi_n^{+}}$ and $\ket{\phi_n^{-}}$ are orthonormal, the matrix elements of the time evolution operator $\hat{U}=\exp(-i\hat{H}t)$ then read
\begin{equation}\label{ume}
 \begin{split}
 c_n&\equiv\brket{1,n+1}{\,\hat{U}\,}{1,n+1}\\
&=\sin^2\theta_n e^{-iE_n^{+}t}+\cos^2\theta_n e^{-iE_n^{-}t},\\
a_n&\equiv\brket{2,n}{\,\hat{U}\,}{2,n}\\
&=\cos^2\theta_n e^{-iE_n^{+}t}+\sin^2\theta_n e^{-iE_n^{-}t},\\ 
b_n&\equiv \brket{2,n}{\,\hat{U}\,}{1,n+1}=\brket{1,n+1}{\,\hat{U}\,}{2,n}\\
&=\sin\theta_n\cos\theta_n\left(e^{-iE_n^{+}t}-e^{-iE_n^{-}t}\right).
 \end{split}
\end{equation}
This needs to be supplemented with the matrix elements of $\hat{U}$ in the state $\ket{1,0}$. From $H\ket{1,0}=E_1\ket{1,0}$ we have
\begin{equation}\label{cm1}
c_{-1}\equiv \brket{1,0}{\,\hat{U}\,}{1,0}=e^{-iE_1t}=e^{i\Omega t/2}.
\end{equation}
By using $\sin(2\theta_n)=\gamma\sqrt{n+1}/\beta_n$, the transition probability for an atom to emit a photon at time $t$, when there are $n$ of them in the cavity, can be expressed as
\begin{equation}
 |b_n|^2=\dfrac{(n+1)\gamma^2}{2\beta_n^2}[1-\cos(2\beta_n t)],
\end{equation}
which displays the Rabi oscillations \cite{ShoreKnight} with the frequency $2\beta_n$ corresponding to the energy gap between the \emph{dressed} atomic states \cite{HarocheR}, determined from \eqref{epm}.

We focus now on the evolution of the density matrix of the total system from an initial factorized state
\begin{equation}
 \hat{\rho}^t=\hat{U}\,\hat{\rho}^0\,\hat{U}^{\dagger},\hspace{0.3cm}\textrm{with}\hspace{0.3cm}\hat{\rho}^0=\hat{\varrho}\otimes\hat{\sigma},
\end{equation}
where $\hat{\varrho}\equiv\hat{\rho}_f^0$ is the initial reduced density matrix of the field and $\hat{\sigma}\equiv\hat{\rho}_a^0$ that for the atoms, and consider the matrix elements of the reduced density matrix of the field at later times
\begin{equation}\label{rhof}
 \brket{n}{\,\hat{\rho}_f^t\,}{n'}=\sum_i\brket{i,n}{\,\hat{\rho}^t\,}{i,n'},
\end{equation}
In terms of the matrix elements of the time evolution operator in \eqref{ume}, we have
\begin{multline}\label{merf}
  \brket{n}{\,\hat{\rho}_f^t\,}{n'}\\= \sigma_{11}\Big[\,c_{n-1}c_{n'-1}^{*}\,\varrho_{n,n'}+b_{n}b_{n'}^{*}\,\varrho_{n+1,n'+1}\,\Big]\\
+\sigma_{12}\Big[\,c_{n-1}b_{n'-1}^{*}\,\varrho_{n,n'-1}+b_{n}a_{n'}^{*}\,\varrho_{n+1,n'}\,\Big]\\
+\sigma_{21}\Big[\,b_{n-1}c_{n'-1}^{*}\,\varrho_{n-1,n'}+a_{n}b_{n'}^{*}\,\varrho_{n,n'+1}\,\Big]\\
+\sigma_{22}\Big[\,b_{n-1}b_{n'-1}^{*}\,\varrho_{n-1,n'-1}+a_{n}a_{n'}^{*}\,\varrho_{n,n'}\,\Big].
 \end{multline}
From this result, it is immediately realized that if the initial reduced density matrices of the subsystems are diagonal, i.e.
\begin{equation}
  \varrho_{n,n'}=\delta_{nn'}P_n^{0},\hspace{0.5cm}\textrm{and}\hspace{0.5cm}\sigma_{ij}=\delta_{ij}\sigma_{ii},
\end{equation}
then the subsequent reduced density matrices remain diagonal (we show this for the atoms next). In this case, with $|a_n|^2=|c_n|^2=1-|b_n|^2$, we have for the probability distribution $P_n^t= \brket{n}{\,\hat{\rho}_f^t\,}{n}$ of the $n$-photon states
\begin{multline}\label{Pnt}
 P_n^t=P_n^0+|b_{n-1}|^2\left(\sigma_{22}P_{n-1}^0-\sigma_{11}P_{n}^0\right)\\-|b_{n}|^2\left(\sigma_{22}P_{n}^0-\sigma_{11}P_{n+1}^0\right).
\end{multline}

For the atoms, we need to trace now over the degrees of freedom of the field. Since we are interested in initial diagonal ensembles for the subsystems, we have
\begin{equation}
\brket{i}{\,\hat{\rho}_a^t\,}{j}=\sum_n\brket{i,n}{\,\hat{\rho}^t\,}{j,n}=K_{11}^{ij}\,\sigma_{11}+K_{22}^{ij}\,\sigma_{22},
\end{equation}
where we have defined
\begin{equation}
 K_{i'i'}^{ij}=\sum_{nn'}\brket{i,n}{\,\hat{U}\,}{i',n'}\brket{i',n'}{\,\hat{U}^{\dagger}\,}{j,n}\,P_{n'}^{0}.
\end{equation}
In terms of the matrix elements in \eqref{ume}, it is easily seen that $K_{i'i'}^{ij}=0$ for $i\neq j$, which implies that the reduced density matrix of the atom also remains diagonal and then, denoting with $p_i^t=\brket{i}{\,\hat{\rho}_a^t\,}{i}$ the occupation probabilities of the atomic states, we have
\begin{equation}\label{pt}
\begin{split}
 p_1^t&=\sigma_{11}+\sum_n |b_n|^2(\sigma_{22}P_n^0-\sigma_{11}P_{n+1}^0),\\
 p_2^t&=\sigma_{22}-\sum_n |b_n|^2(\sigma_{22}P_n^0-\sigma_{11}P_{n+1}^0).
\end{split}
\end{equation}
\section{The one-atom maser}\label{s1m}
So far we have discussed the statistical dynamics of an atom coupled to a single mode of radiation in a cavity. We are interested, however, in the configuration in which the $(N-1)$th atom leaves the cavity when the $N$th one enters it, which is a possible running mode of a one-atom maser, provided the atomic flux through the cavity is adjusted such that each atom is made to spend a time $\tau$ within the cavity. 

\begin{figure}
 \centering
 \includegraphics[scale=0.31]{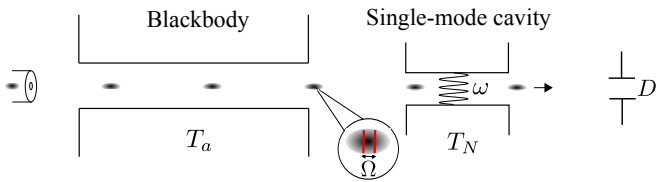}
 \caption{Sketch of a set up for the Jaynes-Cummings thought experiment. After two-level atoms are prepared in arbitrary initial states and made to interact \emph{weakly} with a very large number of photons comprising the stochastic electromagnetic field of a blackbody at temperature $T_a$, the atoms attain, at sufficiently long times, a Boltzmann distribution at temperature $T_a$ \cite{Cummings1}. This is taken as the initial statistical state for the evolution in the presence of a monochromatic field of frequency $\omega$ in the second cavity, which can be assigned an effective temperature $T_N$ at the moment when the $N$th atom enters. The state of the atoms (and hence of the field) after leaving the cavity is probed with a detector $D$.}
 \label{maser}
\end{figure}

In Fig. \ref{maser} we show a set up of the Jaynes-Cumming thought experiment that we have designed for the present discussion. The first cavity on the left is used to prepare two-level atoms with a Boltzmann distribution of their states at temperature $T_a$ after which they are sent to a second cavity to \emph{weakly} interact with a single mode of radiation with frequency $\omega$. 

Since the reduced density matrix of the field remains diagonal for all times, we assume that an effective temperature $T_N$ can be defined as that which makes the probability distribution of the field in the second cavity when the $N$th atom is just entering, a Gibbs distribution at temperature $T_N$. The validity of this depends on the weak coupling condition \eqref{gmG}, as fully discussed in appendix \ref{tpn}.

The only difference with the discussion in the previous section is that each atom now sees a different initial state for the field in the second cavity, which is the final state left by the previous atom so that, denoting
\begin{equation*}
\begin{split}
 P_n^t(N)&: \textrm{probability that the second cavity has $n$ }\\
&\ \;\;\textrm{photons at time $t$ when the $N$th atom is}\\
&\ \;\;\textrm{traversing it,}
\end{split}
\end{equation*}
where $t\in[0,\tau]$, we can rewrite \eqref{Pnt} as
\begin{multline}\label{Pnt2}
 P_n^t(N)=P_n^0(N)+|b_{n-1}|^2\left[\sigma_{22}P_{n-1}^0(N)-\sigma_{11}P_{n}^0(N)\right]\\-|b_{n}|^2\left[\sigma_{22}P_{n}^0(N)-\sigma_{11}P_{n+1}^0(N)\right],
\end{multline}
where $P_n^0(N)=P_n^\tau(N-1)$ is a thermal distribution at temperature $T_N$. Likewise \eqref{pt} is rewritten as
\begin{multline}\label{pt2}
 p_i^t(N)=\sigma_{ii}+(-1)^{i+1}\sum_n |b_n|^2\\
 \times \left[\sigma_{22}P_n^0(N)-\sigma_{11}P_{n+1}^0(N)\right],
\end{multline}
which is \emph{initialized} to $\sigma_{ii}$ every time a new atom enters.

A more suggestive description is however obtained when these equations are differentiated with respect to time. Introducing the notation for the conditional transition probability rates of the $n$-photon states
\begin{equation}\label{rn}
 \dfrac{d}{dt}|b_n|^2=r_{n,n+1}\stackrel{\eqref{ume}}{=}r_{n+1,n},
\end{equation}
we have for the transition probability rates of loosing or gaining a photon in the second cavity, respectively, from the time a new atom enters up to the time $t$
\begin{equation}\label{ws}
\begin{split}
w_{n+1,n}&=\sigma_{11}\,r_{n+1,n},\\w_{n,n+1}&=\sigma_{22}\,r_{n,n+1},
\end{split}
\end{equation}
with $w_{n,n'}=0$ for $|n-n'|>1$ and $n=n'$. In this way \eqref{Pnt2} becomes, after differentiation
\begin{equation}\label{dP}
 \dfrac{dP_n^t(N)}{dt}= \sum_{n'}\left[P_{n'}^0(N)\,w_{n',n}-P_{n}^0(N)\,w_{n,n'}\right].
\end{equation}
Similarly, we define the atomic transition probability rates from the time a new atom enters the second cavity up to time $t$ as
\begin{equation}\label{wa}
\begin{split}
v_{2,1}(N)&=\sum_nP_n^0(N)\,r_{n,n+1},\\ 
v_{1,2}(N)&=\sum_nP_{n+1}^0(N)\,r_{n+1,n},
\end{split}
\end{equation}
with $v_{i,i}=0$. In terms of these, \eqref{pt2} becomes after differentiation
\begin{equation}\label{dp}
\dfrac{dp_i^t(N)}{dt}=\sum_{j} \left[\sigma_{jj}\,v_{j,i}(N)-\sigma_{ii}\,v_{i,j}(N)\right].
\end{equation}
Equations \eqref{dP} and \eqref{dp}, together with the conditions $P_n^0(N)=P_n^\tau(N-1)$ and $p_i^0(N)=\sigma_{ii}$, are recognized as rate equations for a Markov process. A necessary and sufficient condition for the steady state of the field in the second cavity, for which
\begin{equation}\label{Pnss}
 P_n^t(N)=P_n^0(N)\equiv \mathcal{P}_n,
\end{equation}
when many atoms are sent to it ($N\rightarrow\infty$) is that \emph{detailed balance} is satisfied, that is, from \eqref{dP}
\begin{equation}\label{detb}
 \mathcal{P}_{n'}\,w_{n',n}-\mathcal{P}_{n}\,w_{n,n'}=0,\hspace{0.5cm}\forall\,n,n'.
\end{equation}
This is also seen directly from \eqref{Pnt2} by demanding that the quantity
\begin{equation}\label{cd}
 |b_{n}|^2\left(\sigma_{22}\mathcal{P}_{n}-\sigma_{11}\mathcal{P}_{n+1}\right)
\end{equation}
be independent of $n$, so that the second and third terms in \eqref{Pnt2} cancel each other out. For this quantity to be independent of $n$, it is necessary that
\begin{equation}
 |b_{n}|^2\left(\sigma_{22}\mathcal{P}_{n}-\sigma_{11}\mathcal{P}_{n+1}\right)=|b_{\infty}|^2\left(\sigma_{22}\mathcal{P}_{\infty}-\sigma_{11}\mathcal{P}_{\infty+1}\right),
\end{equation}
however, since $\sum_n^{\infty}\mathcal{P}_n=1$, we must have $\mathcal{P}_{\infty}=0$, and then the only way \eqref{cd} is independent of $n$ is by vanishing, giving the condition ($|b_n|^2\leq1$ and only vanishes for isolated values of $n$)
\begin{equation}\label{cond}
 \dfrac{\mathcal{P}_n}{\mathcal{P}_{n+1}}=\dfrac{\sigma_{11}}{\sigma_{22}},
\end{equation}
which is just \eqref{detb} after using \eqref{rn} and \eqref{ws}. It is clearly seen from \eqref{pt2} that when the field in the second cavity reaches a steady state, the probability distribution of the subsequent atoms is unaltered by crossing the second cavity, since each term in the sum in \eqref{pt2} is of the form \eqref{cd}, which was shown to vanish for all $n$.

Since the atoms leave the first cavity with a Boltzmann distribution of their states at temperature $T_a$, we have $\sigma_{22}/\sigma_{11}=e^{-\Omega/T_a}$ and then, if an effective temperature $T_f$ is to be assigned to the steady state of the field, so that $\mathcal{P}_{n+1}/\mathcal{P}_n= e^{-\omega/T_f}$, Eq. \eqref{cond} implies that 
\begin{equation}\label{jcr}
 \dfrac{\omega}{T_f}=\dfrac{\Omega}{T_a},
\end{equation}
which is the same as \eqref{Tf}, that is, only at resonance the two subsystems equilibrate their temperatures.
\section{Entropy production}\label{s3}
\subsection{General considerations}
We have discussed in \cite{Solano2} that, just as a hermitian operator $\hat{A}$ is assigned to every observable in quantum mechanics, thermodynamic observables are also represented by operators which vary sufficiently slow. These are obtained by the projection $\hat{\mathcal{A}}=\mathcal{D}\hat{A}$ of the quantum observable $\hat{A}$ to the Hilbert space spanned by the stationary states $\lbrace\ket{\bm{\alpha}}\rbrace$ of the Hamiltonian $\hat{\mathcal{H}}$ representing the energy of \emph{independent} degrees of freedom (uncoupled subsystems), with $\mathcal{D}$ being the corresponding projection operator. 

For every isolated system, the Hamiltonian can be written as $\hat{H}=\hat{\mathcal{H}}+\hat{V}$, with $\hat{V}$ being the potential mixing all (or a set of) the degrees of freedom left uncoupled by $\hat{\mathcal{H}}$. The evolution of the state, represented by the density matrix $\hat{\rho}_t$, is unitarily generated by $\hat{H}$ and the quantum observable corresponding to entropy is $\hat{S}_t=-\ln \hat{\rho}_t$. 

According to our definition, the part which is observed in thermodynamic phenomena is $\hat{\mathcal{S}}_t=\mathcal{D}\hat{S}_t$, i.e.
\begin{equation}\label{ten}
 \hat{\mathcal{S}}_t=-\mathcal{D}\ln\hat{\rho}_t,
\end{equation}
as can be shown in general by taking its expectation value, which we call the thermodynamic entropy, and noting that it rigorously satisfies the three laws of thermodynamics.

The rate of change of the thermodynamic entropy operator satisfies the equation
\begin{equation}\label{me}
i\partial_t \hat{\mathcal{S}}_t=\mathcal{D}L\hat{\mathcal{S}}_t+ \mathcal{D}Le^{-it\mathcal{N}L}\mathcal{N}\hat{S}_0-i\int_0^td\tau K_\tau\hat{\mathcal{S}}_{t-\tau},
\end{equation}
where $\mathcal{N}=1-\mathcal{D}$ projects operators to their nondiagonal parts, $L=[\hat{H},\cdot]$ is the Liouville superoperator corresponding to $\hat{H}$ and $K_\tau=\mathcal{D}e^{-it\mathcal{N}L}L\mathcal{N}L$ is known as the memory kernel. We use $\partial_t$ in this section as a short-hand notation for $\partial/\partial t$.

For initial states of the local equilibrium form (which are diagonal) and for very weak coupling of the subsystems, the long-time evolution is Markovian, i.e. \eqref{me} becomes memoryless
\begin{equation}
 i\partial_t \hat{\mathcal{S}}_t=-\lim_{s\rightarrow0^{+}}K_s\,\hat{\mathcal{S}}_t,
\end{equation}
with $K_s$ being the Laplace transform of $K_\tau$. The expectation value of this equation leads to
\begin{equation}\label{bdS}
 \avg{\partial_t\hat{\mathcal{S}}_t}=\sum_{\bm{\alpha}\bm{\alpha}'}P_{\bm{\alpha}}W_{\bm{\alpha}\bm{\alpha}'}\ln\frac{P_{\bm{\alpha}}}{P_{\bm{\alpha}'}},
\end{equation}
with the transition rates $W_{\bm{\alpha}\bm{\alpha}'}=2\pi\delta(\varepsilon_{\bm{\alpha}}-\varepsilon_{\bm{\alpha}'})|V_{\bm{\alpha}\bm{\alpha}'}|^2$, calculated in the lowest order in the coupling potential and $P_{\bm{\alpha}}=\brket{\bm{\alpha}}{\hat{\rho}_t}{\bm{\alpha}}$ being the occupation probability of the state $\ket{\bm{\alpha}}$ in its lowest-order approximation, satisfying the transport or Pauli master equation
\begin{equation}\label{pau}
 \partial_t P_{\bm{\alpha}}= \sum_{\bm{\alpha}'}(P_{\bm{\alpha}'}W_{\bm{\alpha}'\bm{\alpha}}-P_{\bm{\alpha}}W_{\bm{\alpha}\bm{\alpha}'}). 
\end{equation}

Eq. \eqref{bdS}, which can alternatively be derived by writing the thermodynamic entropy in this approximation as
\begin{equation}\label{st}
\avg{\hat{\mathcal{S}}_t}=-\sum_{\bm{\alpha}}P_{\bm{\alpha}}\ln P_{\bm{\alpha}}, 
\end{equation}
differentiating and using \eqref{pau}, has the same form as the entropy production of an aged \emph{classical} Markovian system with gaussian fluctuations, where the $\bm{\alpha}$ is in this case interpreted as a realization of the deviation of the thermodynamic (extensive) variables from their equilibrium values.

A special situation occurs when the coupling between subsystems is so weak that they can be treated as statistically uncorrelated as a first approximation. In this case, the thermodynamic entropy becomes additive and an equation like \eqref{pau} holds for the diagonal entries of the reduced density matrix of each subsystem. 

Very weak coupling implies that equilibration within each subsystem takes places much faster than among them and hence, if left alone, the long-time state of the system is one of local equilibrium, described by the factorized density matrix
\begin{equation}\label{ler}
 \hat{\varrho}^{\textrm{r}}=\bigotimes_l\hat{\varrho}_l^{\textrm{r}}= \bigotimes_{l} \exp[-(\hat{\mathcal{H}}_l-\mu_l \hat{\mathcal{N}}_l-\Omega_l)/T_l],
\end{equation}
with $l$ labeling the different uncorrelated subsystems, with temperature $T_l$, chemical potential $\mu_l$ and thermodynamic potential $\Omega_l=\Omega_l(T_l,\mu_l,\left\lbrace x_{\lambda}^l\right\rbrace)$; the latter defined through the normalization of the density matrix, requiring   $\tm{Tr}\, \exp[-(\hat{\mathcal{H}}_l-\mu_l \hat{\mathcal{N}}_l)/T_l]=\exp(-\Omega_l/T_l)$. 

The quantities  $\left\lbrace x_{\lambda}^l\right\rbrace$ are a set of external parameters for subsystem $l$ upon which its Hamiltonian $\hat{\mathcal{H}}_l$ depends; these parameterize the action of the operator which couples the degrees of freedom of the subsystem $l$ to those of the other subsystems, and $\hat{\mathcal{N}}_l$ is the operator corresponding to the number of particles.

Since \eqref{ler} is expressed in terms of (diagonal) thermodynamic operators, the thermodynamic entropy operator from \eqref{ten} and \eqref{ler} is
\begin{equation}\label{Sr}
 \hat{\mathcal{S}}^{\textrm{r}}=-\ln \hat{\varrho}^{\textrm{r}}=\dfrac{1}{T_l}\sum_l(\hat{\mathcal{H}}_l-\mu_l \hat{\mathcal{N}}_l-\Omega_l).
\end{equation}
The equation corresponding to \eqref{bdS} is easily shown to be the sum over all subsystems of the contributions $\avg{d\hat{\mathcal{S}}_l^{\textrm{r}}}$ obtained from
\begin{equation}\label{law1}
T_l\,\avg{d\hat{\mathcal{S}}_l^{\textrm{r}}}=\avg{d\hat{\mathcal{H}}_l}-\mu_l\avg{d\hat{\mathcal{N}}_l}+\sum_{\lambda}F_{\lambda}^ldx_{\lambda}^l,
\end{equation}
where $F_{\lambda}^l=-\avg{\partial \hat{\mathcal{H}}_l/\partial x_{\lambda}^l}$. This may be written as the well-known form of the first law of thermodynamics
\begin{equation}\label{tds}
T_l\,\partial_t\avg{\hat{\mathcal{S}}_l^{\textrm{r}}}=\partial_t\avg{\hat{\mathcal{H}}_l}-\mu_l\,\partial_t\avg{\hat{\mathcal{N}}_l}+\sum_{\lambda}F_{\lambda}^l\,\partial_t x_{\lambda}^l, 
\end{equation}
provided the process takes place slowly enough, a condition which can be stated as $\lVert\partial_t\hat{\varrho}^{\textrm{r}}\rVert$ being very small, in a norm defined in \cite{Solano2}.

When the continuity equations for the local average energy and number of particles is used in \eqref{tds}, an expression for the total entropy production $\Pi=\partial_t\avg{\hat{\mathcal{S}}^{\textrm{r}}}$ in the system can then be obtained. This will be done for the particular case of two subsystems in contact in the following section.

\subsection{Application to the one-atom maser}
The dynamics of the Jaynes-Cummings model is special because, in the  basis $\lbrace\ket{\bm{\alpha}}\rbrace=\lbrace\ket{1,n+1},\ket{2,n}\rbrace$, the diagonal entries of the reduced density matrices of each subsystem undergoes Markovian evolution (as in \eqref{pau}) independent of the strength of the coupling $\gamma$, as expressed by equations \eqref{dP} and \eqref{dp}. 

The very weak coupling condition enters the calculation of the thermodynamic entropy when correlations between the field and atoms are considered negligible to a first approximation so that \eqref{st} is additive among subsystems
\begin{equation}
\mathcal{S}_N^t=-\sum_i p_i^t(N)\ln p_i^t(N)-\sum_n P_n^t(N)\ln P_n^t(N). 
\end{equation}
After using the conservation of probabilities, we then have for the entropy production $\Pi_N^t=d\mathcal{S}_N^t/dt$
\begin{equation}\label{PiN}
 \Pi_N^t=-\sum_i \dfrac{dp_i^t(N)}{dt}\,\ln p_i^t(N)-\sum_n \dfrac{dP_n^t(N)}{dt}\,\ln P_n^t(N).
\end{equation}
Since the state of the field is monitored through the atoms when these just leave the second cavity, we calculate \eqref{PiN} at time $t=\tau$ by using \eqref{dP} and \eqref{dp}
\begin{equation}\label{PiN2}
 \Pi_{N}^\tau = \left(\dfrac{\Omega}{T_a}-\dfrac{\omega}{T_{N+1}}\right)J_N^P,
\end{equation}
where $J_N^P$ is the photon current from the field to the $N$th atom when this is just leaving the second cavity
\begin{equation}\label{JNP}
 J_N^P=\sum_n\left[P_{n+1}^0(N)\,w_{n+1,n}-P_{n}^0(N)\,w_{n,n+1}\right],
\end{equation}
with the $w$'s evaluated at time $t=\tau$. In arriving at this result, we have used the fact that $p_i^\tau$ is initialized to $p_i^0=\sigma_{ii}$, by construction, and that the state of the field each time a new atom enters the cavity can be assigned an effective temperature $T_N$, i.e.
\begin{equation}
P_{n+1}^0(N)/P_{n}^0(N)=e^{-\omega/T_N}. 
\end{equation}
This corresponds to the local equilibrium state discussed in \eqref{ler} here evolving under a discrete dynamics: every time a new atom enters the cavity, both the atom and the photon field are thermal. 

The discussion in the previous section then suggests that the entropy production calculated in \eqref{PiN2} should be directly connected to that obtained using the arguments after \eqref{tds}.  To this end, consider two classical subsystems which are in contact at temperatures $T_1$ and $T_2$ and respective chemical potentials $\mu_1$ and $\mu_2$. The total entropy production $\Pi$ for this system is
\begin{equation}\label{Pic}
 \Pi = \dfrac{1}{T_1}\dfrac{\delta Q_1}{dt}+\dfrac{1}{T_2}\dfrac{\delta Q_2}{dt},
\end{equation}
where $\delta Q_i$ is the quantity of heat exchanged by the subsystem $i$ with the other subsystem during a time interval $dt$, which is expressed at constant volume (constraints specified by the $x_{\lambda}^i$ in \eqref{tds} kept \emph{fixed}) by
\begin{equation}\label{fl}
 \delta Q_i = dE_i-\mu_i dN_i,
\end{equation}
where $E_i$ and $N_i$ are, respectively, the average internal energy and number of particles of subsystem $i$. Using the conservation of energy and particles we have, in terms of the energy current $J^E$ and a particle current $J^P$ 
\begin{equation}\label{epc}
J^E=\dfrac{dE_1}{dt}=-\dfrac{dE_2}{dt},\hspace{0.5cm} J^P=\dfrac{dN_1}{dt}=-\dfrac{dN_2}{dt}.
\end{equation}
Substituting \eqref{fl} and \eqref{epc} in \eqref{Pic}, we can write the total entropy production as the sum $\Pi=\Pi^E+\Pi^P$ of the contributions from the two irreversible processes taking place: energy transport and particle diffusion, with the respective terms given by
\begin{equation}\label{Pic2}
 \Pi=\left(\dfrac{1}{T_1}-\dfrac{1}{T_2}\right)J^E+\left(\dfrac{\mu_2}{T_2}-\dfrac{\mu_1}{T_1}\right)J^P.
\end{equation}

Thermodynamic equilibrium between the subsystems is characterized by the vanishing of the thermodynamic forces (quantities in parentheses in \eqref{Pic2}) and the respective induced currents, and therefore of the total entropy production. This leads to the well-known conditions for equilibrium $T_1=T_2$ and $\mu_1=\mu_2$.

Having in mind these considerations, we observe that when the steady state of the field in the second cavity is approached, 
\begin{equation}
P_n^0(N)\rightarrow \mathcal{P}_n\hspace{0.3cm}\therefore\hspace{0.3cm}T_N\rightarrow T_f. 
\end{equation}
By using \eqref{jcr} and the detailed balance condition \eqref{detb} in \eqref{PiN2} and \eqref{JNP} we then see that both the thermodynamic force (quantity in parentheses in \eqref{PiN2}) and the induced current vanish in the steady state and hence the total entropy production. This shows that the steady state is a state of thermodynamic equilibrium. 

An even more interesting insight is obtained when the nature of the transport processes with which equilibrium is approached in the second cavity is inquired. Since the particles being transported between the atoms and the field are photons, at resonance when $\omega=\Omega$ the energy current is $J^E=\omega J^P=\Omega J^P$ and then in \eqref{PiN2} we have $\Pi_N^\tau\rightarrow(1/T_a-1/T_f)J^E$, which is of the form $\Pi^E$ in \eqref{Pic2}. Out of resonance, when $\omega\neq\Omega$, we see instead that $\Pi_N^\tau\rightarrow(\Omega/T_a-\omega/T_f)J^P$, which is of the form $\Pi^P$ in \eqref{Pic2}, with $-\omega$ and $-\Omega$ playing the role of chemical potentials for the field and atoms, respectively.

We note that in obtaining the above description of the problem, we have pictured the dressed photons as having a chemical potential which is different (when $\omega\neq\Omega$) depending on whether these belong to the cavity field or to the atoms. A photon ``belonging'' to an atom is thought as absorbed by the atom, with the potential of giving it back to the field. 

This identification of the chemical potential of the photons in the cavity, interacting with the atoms, requires abandoning the common belief that a vanishing chemical potential is a property of all photons (as in a black-body field), which is not generally so \cite{Wurfel}. The chemical potential of radiation depends on the emitter and can even be positive if the latter has an energy gap through which the radiative transitions occur. 

The remarks above therefore provide us with an interpretation of the results of the Jaynes-Cummings thought experiment. At resonance, energy transport is active when approaching the steady state and particle diffusion is absent (impermeable wall building up between the subsystems), a reason why the temperatures of the atoms and field equilibrate. Out of resonance only particle diffusion is active and energy transport is absent (adiabatic wall building up between the subsystems) and then the temperatures do not equilibrate. 

The interference of different irreversible processes taking place in a thermodynamic system is basically a classical phenomenon formalized long ago by Onsager \cite{Onsager,Onsager2}. The fact that in the present situation, and under our interpretation, the processes become mutually exclusive depending on the detuning conditions of the cavity is a fully quantum-mechanical phenomenon: we are dealing with single atoms and photons.

\section{Conclusion}\label{s4}
By calculating the quantum entropy production in a system of atoms weakly interacting with a cavity field and comparing it to the classical results expected for a Markovian evolution, we have been able to infer the nature of the processes with which the steady state in a one-atom maser is approached both at resonance and out of resonance. In particular, we have shown that the lack of equilibration of temperatures of the subsystems in the off-resonant situation is consistent with the equilibrium state attained by two subsystems in contact but adiabatically insulated from each other. At resonance, this insulation is not present and therefore the temperatures do equilibrate. The results presented here are representative of the new kind of information which can be gained by applying quantum-thermodynamic methods.

\section{acknowledgments}
I would like to acknowledge the support from the Fulbright-Colciencias Fellowship and the Columbia GSAS Faculty Fellowship, as well as helpful discussions with Prof. Andrew Millis, Junichi Okamoto and Zhouran He.

\appendix
\section{Thermal distribution of the photon field when atoms enter the cavity}\label{tpn}
The condition of weak coupling in \eqref{gmG} is fundamental for the present discussion since only in this case a truly thermodynamic behavior corresponds to the quantum statistics. This is actually realized in practice in cavity QED experiments in which $\gamma/\Omega\sim 10^{-7}$ is easily obtained \cite{Brune2}, with very long decay times for the cavity $\tau_{\tm{cav}}/\tau \sim 10^{4}$ and the atomic states $\tau_{\tm{atom}}/\tau \sim 10^{3}$ \cite{Walther}. 

We assume that it is impossible to have an infinite number of photons in the cavity at any time (e.g. that costs a lot if the photons have nonzero chemical potential) or, equivalently, that an integer $M$ should exist such that, if there are $n$ photons in the cavity, $n\le M$. The weak very coupling condition can then be expressed as
\begin{equation}\label{wcc}
 0<\gamma \tau \ll M^{-1}.
\end{equation}
By expanding the transition probability $|b_n|^2$ evaluated at time $\tau$ in a power series in $\gamma \tau$ we then find, up to the leading order,
\begin{equation}\label{bnwc}
 |b_n^\tau|^2=(n+1)(\gamma\tau)^2.
\end{equation}
If the probability distribution when the $N$th atom is entering the cavity is assumed to be thermal at temperature $T_N$ we should have
\begin{equation}\label{Pkn}
 P_n^0(N)=a_N K_N^n, \hspace{0.5cm}K_N=e^{-\omega/T_N},
\end{equation}
where $a_N=1-K_N$ is the normalization constant. Note that the factor coming from the chemical potential $\mu$ is absorbed in the definition of $a_N$, i.e. $a_N\propto\,e^{\mu/T_N}$.

When \eqref{Pkn} is substituted in \eqref{Pnt2} at time $\tau$ we have
\begin{multline}
P_n^{\tau}(N)=a_N K_N^n\Bigl[1+K_N^{-1}(\sigma_{22}-K_N \sigma_{11})\\\times(|b_{n-1}^\tau|^2-K_N |b_n^\tau|^2)\Bigr]. 
\end{multline}
The condition $P_n^{\tau}(N)=P_n^{0}(N+1)$ together with \eqref{Pkn} imply that we must have 
\begin{multline}\label{kn1}
a_{N+1} K_{N+1}^n=a_N K_N^n\Bigl[1+K_N^{-1}(\sigma_{22}-K_N \sigma_{11})\\\times(|b_{n-1}^\tau|^2-K_N |b_n^\tau|^2)\Bigr]. 
\end{multline}
In order to satisfy this equation it is required that
\begin{multline}\label{dtn}
e^{\frac{n\omega (\Delta T_N/T_N)}{T_N+\Delta T_N}}(1+X_N)=1+K_N^{-1}(\sigma_{22}-K_N \sigma_{11})\\\times(|b_{n-1}^\tau|^2-K_N |b_n^\tau|^2). 
\end{multline}
Replacing this in \eqref{kn1} give the recursive equations
\begin{eqnarray}
 a_{N+1}&=&(1+X_N)\,a_N,\label{aN}\\
 T_{N+1}&=&T_N+\Delta T_N.\label{TN1}
\end{eqnarray}
The possibility of having a thermal distribution for the field every time a new atom enters the cavity then relies on the self-consistency of the equations \eqref{aN} and \eqref{TN1}. 

Note that this description is symmetric with respect to making $\gamma$ very small or instead making $\tau$ very small. In the latter case, it is not expected that the effective temperature of the field changes considerably after the passage of an atom and then $\Delta T_N/T_N\ll1$. 

Assuming $X_N\ll1$ as well, we can then expand the left-hand side of \eqref{dtn} and substitute \eqref{bnwc} in the right-hand side to obtain
\begin{equation}\label{woT}
 (\omega/T_N)(\Delta T_N/T_N)=K_N^{-1}(\sigma_{22}-K_N \sigma_{11})(1-K_N)(\gamma \tau)^2,
\end{equation}
which determines the temperature changes at each step of the discrete dynamics, and
\begin{equation}\label{XN}
 X_N=-(\sigma_{22}-K_N \sigma_{11})(\gamma \tau)^2,
\end{equation}
which keeps the probability distribution of the photons normalized.

Using \eqref{Pkn} we have, in terms of the steady state temperature of the photon field $T_f=(\omega/\Omega)\,T_a$, 
\begin{equation}\label{dTN}
\begin{split}
 \Delta T_N/T_N&= -\,\tm{sech}\Bigl(\dfrac{\omega}{2T_f}\Bigr)\sinh\Bigl[\dfrac{\omega}{2}\Bigl(\dfrac{1}{T_f}-\dfrac{1}{T_N}\Bigr)\Bigr]\,\\
 &\ \hspace{0.3cm}\times\tm{sinhc}\Bigl(\dfrac{\omega}{2T_N}\Bigr)(\gamma \tau)^2.
 \end{split}
\end{equation}
This proves the consistency of \eqref{TN1} since, beginning with a temperature $T_i=T_{N=1}$ for the field, if $T_i>T_f$, \eqref{dTN} says that the field will start cooling as atoms cross the cavity until the temperature of the steady state $T_f$ is reached after which no more temperature changes occur. Heating of the field will happen instead if $T_i<T_f$. 

It should be mentioned, as seen from \eqref{dTN}, that if $\omega/T_f<1$ the cooling process can take place from \emph{any} initial temperature $T_i>T_f$. However, the heating process can not start from arbitrarily low temperatures since the prefactor of $(\gamma\tau)^2$ in \eqref{dTN} can grow arbitrarily, with the possibility of making $\Delta T_N/T_N\ll1$ inconsistent. The same reasoning applied to the case $\omega/T_f>1$ shows that cooling consistently takes place only for sufficiently large initial temperatures. 

These observations are in line with the third law of thermodynamics which, in one of its forms, states that the specific heat of materials tend to zero as the temperature goes to zero or, in other words, it is very hard to change the temperature when the system is close to the absolute zero.

We now turn our attention to the consistency of \eqref{aN}. This is proven by rewriting that equation, on one hand, in terms of $a_N=1-K_N$ and using \eqref{XN}
\begin{equation}\label{KN1}
 K_{N+1}=K_{N}+(\sigma_{22}-K_N\sigma_{11})(1-K_N)(\gamma\tau)^2.
\end{equation}
On the other hand, by expanding $K_{N+1}$ in a Taylor series about $\Delta T_N/T_N=0$ we easily get, up to leading order,
\begin{equation}
 K_{N+1}=K_{N} +K_{N}(\omega/T_N)(\Delta T_N/T_N),
\end{equation}
which, after using \eqref{woT}, is equivalent to \eqref{KN1}. 

We have therefore proven the consistency of both \eqref{aN} and \eqref{TN1} and then the probability distribution of the field every time a new atom enters the cavity is really thermal in the weak coupling limit.

\bibliography{references}

\end{document}